# A Study of S doped ZnSb

X.Song and T.G.Finstad[1]
Department of Physics, P.O. Box 1048 Blindern, N-0316 OSLO, University of Oslo, Norway

ABSTRACT

We report on S-doping of ZnSb for S concentrations ranging from 0.02 at% to 2.5 at%. There are no previous reports on S-doping. ZnSb is a thermoelectric material with some advantages for the temperature range 400 K - 600 K. The solid solubility of S in ZnSb was estimated to be lower than 0.1% from observations of precipitates by scanning microscopy. Hall and Seebeck measurements were performed as a function of temperature from 6K to 623 K. The temperature dependence of the electrical properties suggests that S introduces neutral scattering centers for holes in the p-type material. An increase in hole concentration by S is argued by defect reactions involving Zn vacancies.

---

[1] Corresponding author.

email address: tgf@mail.uio.no (T.G.Finstad)

# 1. Introduction

The existence of the thermoelectric material ZnSb has been known for a very long time [1, 2]. Today we consider ZnSb as a thermoelectric compound with renewed interest [3] and potential for use in the important temperature range 400 K - 600 K where other known efficient thermoelectric materials have problems with toxicity or abundance of the constituent elements. There are currently also other applications of ZnSb actively being explored such as electrodes for rechargeable Li ion batteries [4] or as phase change memory cells [5]. ZnSb has been prepared by different methods [2], yielding polycrystalline material [6-8] or single crystals [9-11]. Undoped, pure ZnSb has always been reported to be a p-type semiconductor. The hole concentration can be controlled by adding various amounts of different dopants. However, to obtain n-type ZnSb has proven to be difficult. This is unfortunate since thermoelectric devices such as power generators or Peltier elements for cooling usually consist of p-type legs that are thermally in-parallel and electrically in-series with n-type legs and thermal mismatch problems are minimized by making both (n-and p-type) legs from the same semiconductor.

The electronic structure and band diagram of ZnSb has recently been studied theoretically by several groups [12-17], and we judge that the conduction band itself has favorable features for a n-type thermoelectric material in terms of density of states and dispersion. The challenge of n-type material is to find donor atoms that can be introduced to a sufficiently high concentration, $N_D$, which for thermoelectric applications is considered to be $N_D \gtrsim 10^{19}$ cm$^{-3}$. Ideally, the doping process should introduce no deep levels in the band gap that could compensate the effect of dopant additions. While theoretical studies might be appealing, we feel that doping needs to be tested by experiments.

The elements Ag, Au, Sn, Pb [11, 18] and Cu [14, 19, 20] act as accepters probably by lattice site substitutions such as: $Ag_{Zn}$, $Au_{Zn}$, $Sn_{Sb}$, $Pb_{Sb}$ and $Cu_{Zn}$. Similarly the elements K and Na have been suggested as acceptors by Zn site substitutions ($K_{Zn}$ and $Na_{Zn}$)[21]. There are several negative reports on producing n-type ZnSb. In and Te could be expected to act as donors by substituting for Zn and Sb respectively ($In_{Zn}$ and $Te_{Sb}$) by counting valence electrons. Still In and Te has been reported to increase the hole concentration, increase the electrical conductivity and decrease the Seebeck co-efficient at relatively high concentrations, and some formation of InSb and ZnTe was reported [22]. The group 13 element In has been found to act as a donor in CdSb [22], which has the same crystal structure as ZnSb. It is then probably substituting Cd as $In_{Cd}$. It was later reported that ZnSb also could be made n-type by In doping, although it was unstable and converted back to p-type [23]. The group 13 element Ga could act as a donor by substituting Zn forming $Ga_{Zn}$. Some degree of compensation might have occurred, but Justi et al. did not achieve n-type ZnSb with Ga despite several attempts with single and polycrystalline ZnSb [24]. Several groups have studied the case of Te addition. Ueda et al. [25] summarizes the efforts when presenting their own work; At low concentrations (<1 at%) and high concentrations (>3 at%) the samples are p-type while in a small range around 2 at% Te gives n-type, possibly by forming substitutional Te atoms on Sb sites: $Te_{Sb}$. Excess doping with Te results in precipitation of the ZnTe phase and a change in conduction from n- to p-type. Niedziolka et al. [17] predicted theoretically by DFT calculations that boron would electronically be a good candidate for n-type ZnSb, but did not succeed to synthesize the material and suggested that the formation energy of a B atom on a Zn site in ZnSb is too high.

We here report a study on S doping of ZnSb. That has not been reported before. It was inspired by the indications of partial success with Te doping to produce n-type ZnSb. Sulfur is a group 16 element as Te and could be expected to substitute Sb to yield n-type. One can learn more about conduction in ZnSb from the results, even if no n-type conduction was observed in the present case. We briefly describe the synthesis procedure in section 2. We produced a series of samples (series A) for initial studies. That synthesis procedure produced inhomogeneous samples, unsuited for detailed analysis of transport measurements, but indicated solubility limits and no n-type. Another synthesis procedure (series B) produced samples suited for detailed transport measurements, which is the main part of the investigation. As a consequence of S doping, an increase in the hole concentration and an increase in neutral scattering centers was observed. These results were discussed and a tentative simple hypothesis involving clusters of point defects could apparently explain the main trends of the measurements.

## 2. Experimental methods

We have produced two series of ZnSb samples doped with various concentrations of sulfur ([S] = 0.02 -2.5 at%). Series A was for initial studies and consisted of samples synthesized by melting the elements in evacuated quartz ampules which were cooled down in air before they were annealed at 623K for 120 hours. We will refer to this series of samples as 'melt solidified samples' or series A.

Another series, series B, was synthesized similar to series A for the first part of the synthesis. The first part was followed by grinding the material into powder and then hot-pressing into pellets with dwelling time of 30 min at 720 K. The cool-down to room temperature took about 24 hours. We will refer to these samples as 'hot-pressed samples' or series B.

For interpretation and discussion of measurement results it is useful to compare with measurements done on the same setup on undoped ZnSb prepared for other purposes. The details of synthesis of those samples have been described before [19].

The pellets were cut into pieces and polished for electrical characterization. We measured the temperature dependence of resistivity, Hall carrier concentration, Hall mobility and Seebeck coefficient. In the temperature range 10K -300 K we used a Lake Shore 7604 system, while in the range 300 K - 600 K we used a MMR based sample chamber with custom electronics. For the Seebeck measurements we used a custom built setup described before [26], while here only the room temperature values are reported.

The structural characteristics of the samples where done by optical microscopy, scanning electron microscopy (SEM, using Hitachi TM 3000) with EDS analysis and X-ray diffraction (XRD, using Bruker D8 DISCOVER with Cu Kα).

## 3. Results and discussion

3.1 Structure characterization series A (melt solidified)

The structure and phase constituents of the samples were studied by combining optical microscopy, SEM with EDS and XRD. Fig.1 gives some typical examples of the surface appearance of samples from series A with different overall nominal S concentrations. The even gray areas are identi-fied as the ZnSb phase One can see that the samples in Fig. 1 a) and 1 b) are very inhomogeneous. The dark bands are attributed to ZnS precipitated phase. The bright color is attributed to Sb phase. The dendrite structures were not observed by microscopy for samples with [S] ≤ 0.1 at%, as indicated by Fig. 1 c) showing a sample with [S]=0.02 at%. There were no indication of segregation of ZnS and the dark contrast is attributed to cracks and pits.

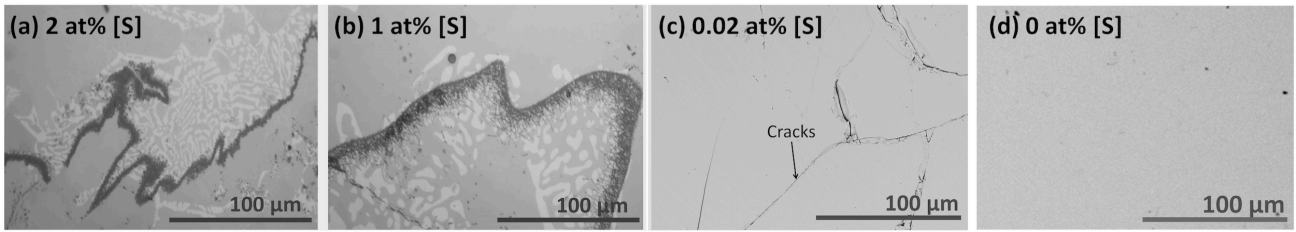

Fig. 1. Images showing typical surfaces of melt solidified ZnSb samples with different S content. a) Optical microscopy for [S]=2 at%, b) Optical microscopy [S]=1 at%, c) SEM micrograph [S]=0.01 at%. d) SEM micrograph [S]=0.01 at%. The long dark bands in a) and b) are ZnS (secondary phase), while the bright dendritic pattern is Sb. In c) only cracks and pits can be identified. The matrix in all the pictures is the ZnSb phase.

Fig. 2 shows XRD patterns from the samples with [S]= 2 at%, and [S]= 0.02 at% respectively. The majority phase in both cases is identified as ZnSb. The secondary phase in the sample doped with [S]= 2 at% was determined to be ZnS while this phase was not found in the sample with [S]= 0.02 at%. A similar difference was found between other samples dependent on [S]: For [S]>0.1 at% we observed clearly ZnS precipitation, while for [S] ≤ 0.1 at% we observed no ZnS precipitation by XRD. This is in agreement with the microscopy study. We therefor deduce that the maximum solid solubility of sulfur in ZnSb is less than 0.1 at% for the conditions sample series A has been through. Analysis of the present situation indicate that the solid solubility is definitely lower than 0.1 at%. To put that in context of possible doping, 0.1 at% corresponds to a sulfur atomic density of $4\times10^{19}$ at/cm$^3$.

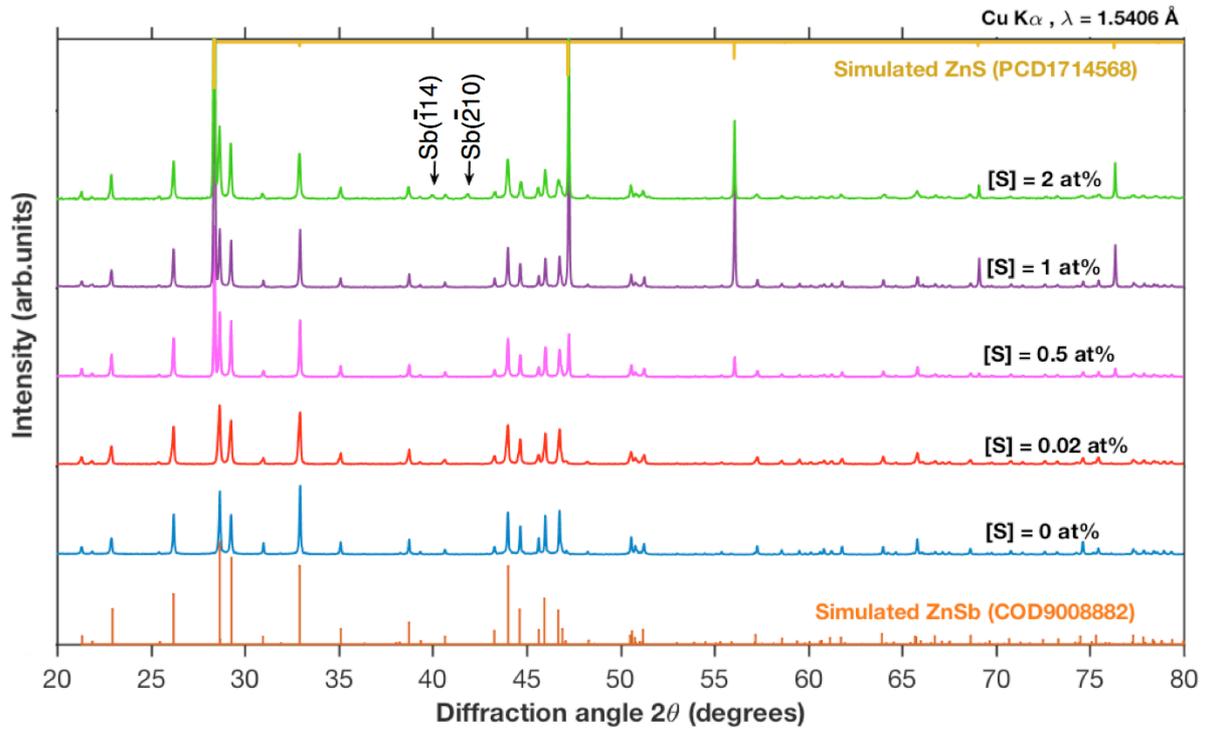

**Fig. 2.** X-ray diffraction patterns of different melt solidified ZnSb samples of series A with different overall sulphur concentrations, [S], marked on each pattern. The calculated (from the XRD database [27]) pattern for ZnSb and ZnS is shown at the bottom and top of the graph, respectively. All peaks are accounted for by identifying the lightly doped as ZnSb, for the heavily doped case there are strong peaks from ZnS and Sb in addition to the peaks from ZnSb. This is consistent with precipitates observed in SEM with EDS.

.

We also have observed Sb-precipitation in the samples, an example is shown for the [S]=2 at% case of Fig. 2. The degree of precipitation depends upon the sulfur concentration, for example in the XRD pattern for the [S]=0.02 at% case of Fig. 2, no Sb peaks can be detected. There are reports about Sb-precipitations in many studies of ZnSb [6, 19]. In the present case of sulfur doping, it is reasonable that there is much Sb precipitation when there is much ZnS formation. The precipitation and ZnS second phase make the material inhomogeneous. This makes the interpretation of electrical measurements doubtful. We believe that the properties of ZnSb will dominate, while great precautions must be exercised in interpreting electrical properties of series A samples, but the measurements could at least provide an indication of expected behavior for more homogenous samples.

3.2. Electrical characterization, Series A (melt solidified)

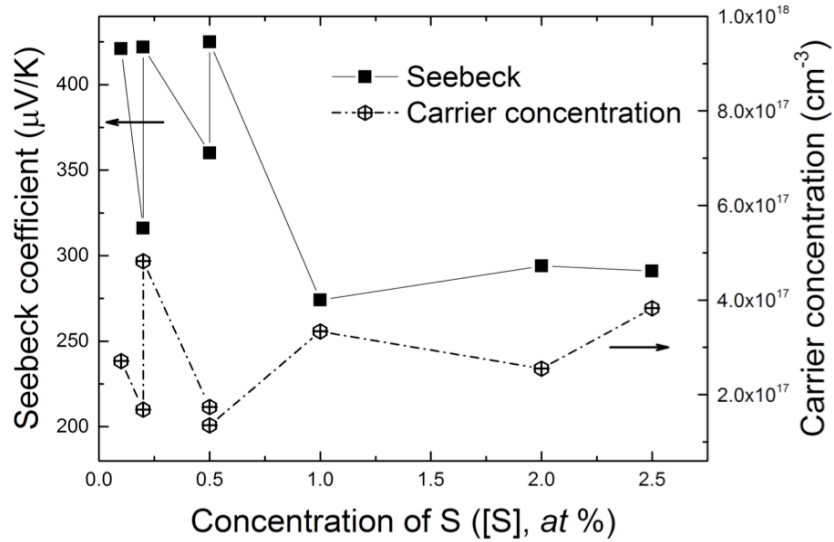

Fig.3. The Seebeck coefficient and the Hall carrier concentration vs. sulfur concentration at room temperature for samples of series A. Different points are from different samples of melt solidified ZnSb with the specified concentration of sulfur added to the melt. These samples are highly nonuniform(see Fig.1).

Fig. 3 shows some results of electrical characterization of the melt-solidified samples of series A with different nominal content of sulfur. Both the Seebeck coefficient and the carrier concentration are shown as a function of the sulfur concentration, [S]. The sign of the Seebeck coefficient and the sign of the Hall voltage indicate that the samples behave as a p-type material. If the electrical properties are dominated by ZnSb, the measurements indicate that S does not yield n-type. Considering the inhomogeneity, most of the details should not be interpreted; however, preparing homogenized material and measuring the temperature dependence of the electrical properties could reveal some of the electronic processes occurring in S doped ZnSb. That is the reason why we will present the results of sample series B.

3.3. Structure characterization series B (hot-pressed powders)

The hot-pressed samples are more homogenized than those of series A, as seen from Fig. 4. Small bright particles are Sb phase precipitates. We have noticed a systematic difference in the amount of Sb phase among the samples. It appears that the general tendency is that the higher the S concentration, the more Sb phase there is. This is similar to what was observed for sample series A, and follows naturally from the reaction between Zn and S to form ZnS, which leaves excess Sb.

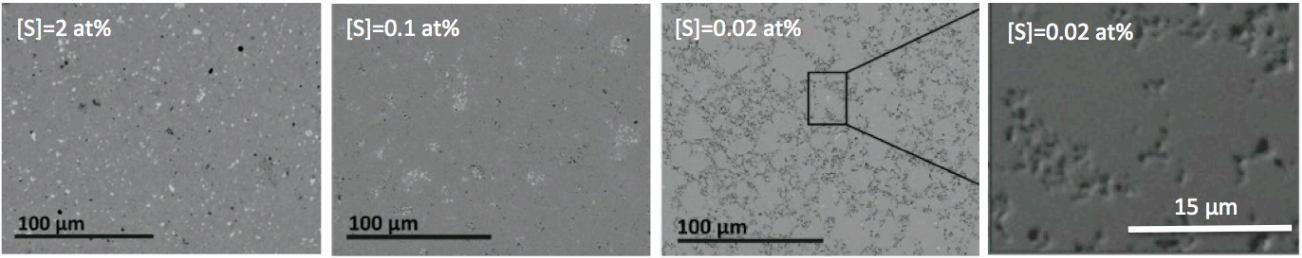

Fig 4. Scanning electron microscopy on hot-pressed ZnSb of series B. The overall sulfur concentration is annotated in the Figs. Small bright spots are small Sb precipitates distributed homogeneously. For [S]= 0.02 at% most of the contrast is attributed to topography of the porous sample.

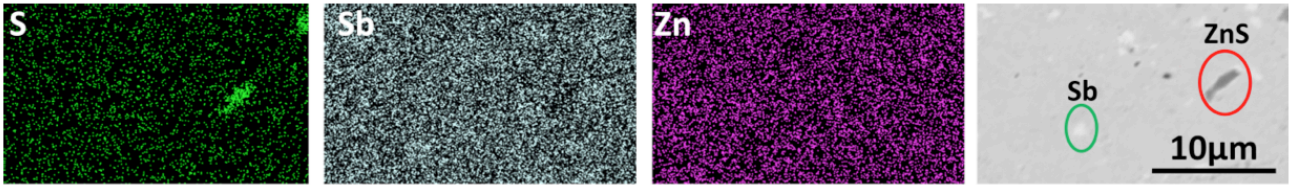

Fig. 5. EDS mapping of elements together with a SEM micrograph on ZnSb image of series B with [S] = 2 at%. The dark precipitations seen in the SEM to the right is ZnS while the bright dots from Sb.

Fig. 5 shows the EDS mapping result for a sample with [S]=2 at%. We notice that there is a short precipitate of ZnS. The continuity of ZnS phase that has been observed in Fig. 1 and Fig. 2 (as lines) has been broken down by ball-milling and when the overall S concentration is above the solid solubility they do not dissolve.

3.4 Electric characterization of series B (hot-pressed samples)

3.4.1. Carrier concentration

Fig. 6 shows the Hall carrier concentrations of hot-pressed samples of series B with [S]=0.02 at% and [S]= 0.1 at% respectively measured between 7 K and 623 K. Measurements of undoped ZnSb are also included for comparisons. The sign of the Hall voltage implies that holes are dominating and no n- type behavior is found for these samples as with the samples of series A. The Hall carrier concentrations for sulfur doping are also higher than that for undoped ZnSb. We also notice from Fig. 6 that there is no 'freeze-out' of the carriers at low temperature for any of the samples, rather the Hall carrier concentration increases with decreasing temperature for the lowest measurement temperatures. For the undoped ZnSb case this has been shown to be associated with so-called impurity band conduction [28]. It is also interesting to note that the curves all have a similar steady increase with temperature in the temperature range 100-300 K. For the undoped case it was shown that this could be associated with the ionization of an acceptor level situated around 0.069 eV above the valence band [28]. The similar behavior for the samples is taken as an indication that the electrical properties of the S doped samples are influenced by intrinsic defect states, where Zn vacancy related defects are the prime suspects.

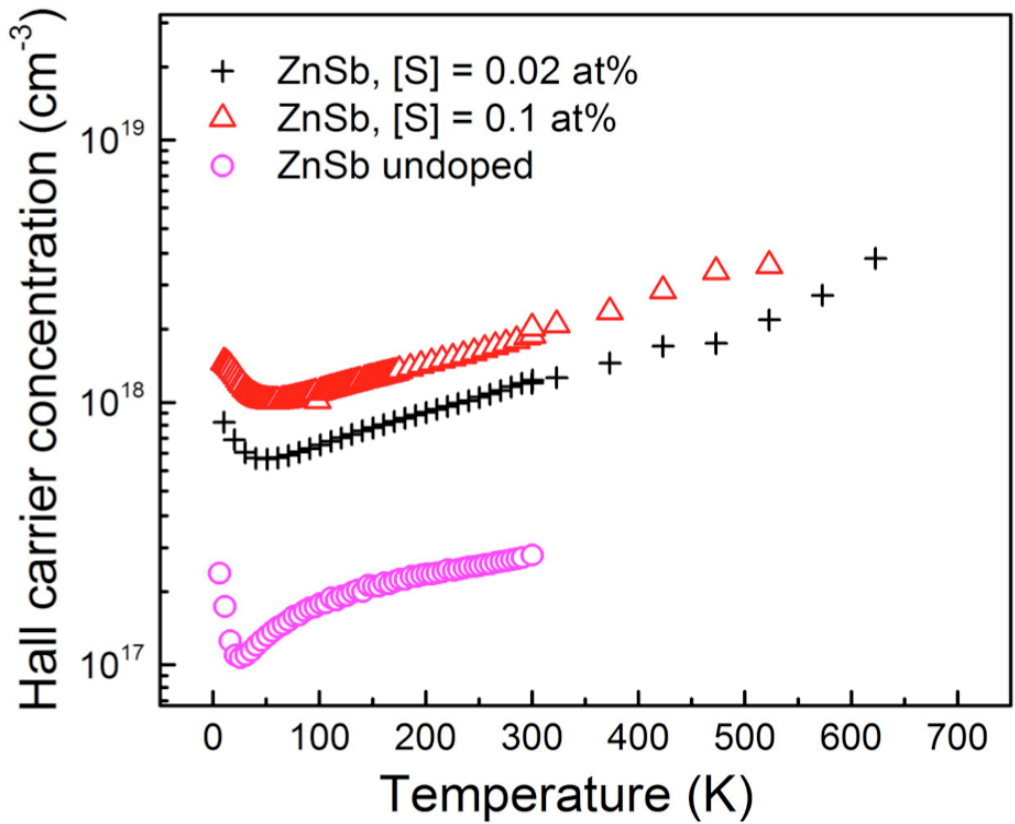

Fig. 6. Hall carrier concentration measurement from 7K to 623 K for hot pressed samples in series B doped with sulfur as indicated. For comparison a sample with undoped ZnSb also shown. The sign of the Hall voltage is positive.

3.4.2. Resistivity

Fig. 7 shows the resistivity as a function of temperature for series B samples. The upper Fig. 7a) contains 2 samples with [S]=0.02 and 0.1 at%, respectively. To avoid clutter in the figure, the curve for undoped ZnSb is shown in a separate plot in Fig 7 b). We see from Fig. 7 that the sign for the temperature coefficient of resistivity, $(\partial\rho/\partial T)/\rho$, is changing with temperature for the undoped ZnSb sample, but is always negative for the sulfur doped samples. Since the temperature variations of the carrier concentrations are similar for these samples (Fig. 6) we need to turn to the temperature dependence of the mobility in order to understand the difference between the samples of Fig. 7. We will return to discussion of the sign further on in the discussion and here just remark that the reason we bring it up is to clear up misunderstandings of the temperature behavior among peers.

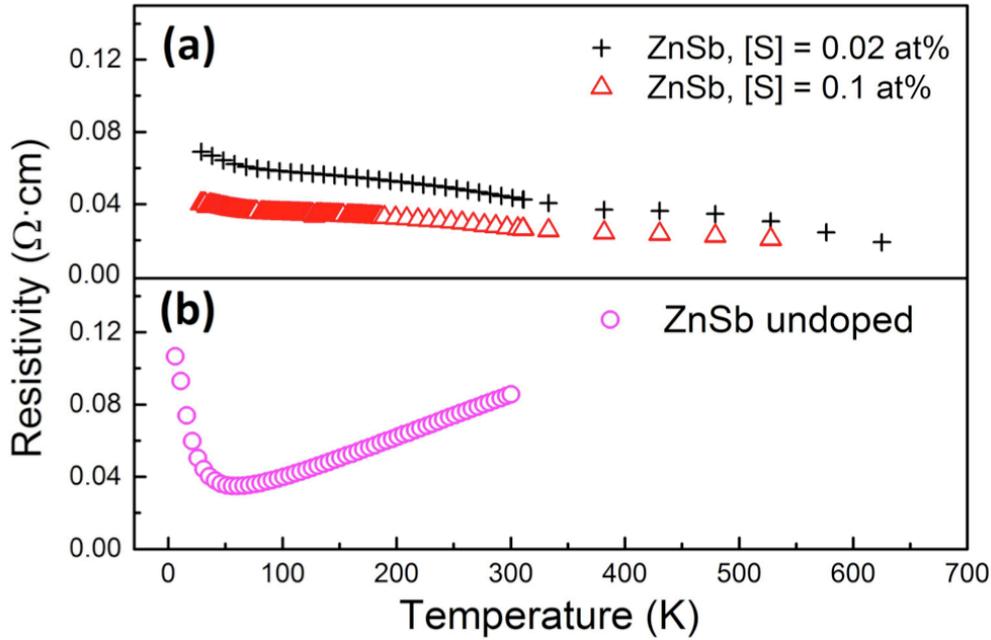

Fig. 7. Resistivity measurement from 7 K to 623 K for a) hot pressed samples in series B doped with sulfur as indicated. b) Undoped ZnSb for comparison.

3.4.3. Mobility

Fig. 8 shows the Hall mobility as a function of temperature. The mobility measured for undoped ZnSb is shown for comparison. It is seen that the mobility of the doped sample is much lower than that of undoped ZnSb. (This is the case for the doped samples omitted from the graph too.) The overall trend when going to the highest temperature is in agreement with the expectation that the lines converge towards a common line where phonon scattering dominates. However the temperature regions around the maximum needs some discussion. The maximum and the sign change of slopes of the curves are usually taken to signal the change in dominating scattering mechanism. The overall shape of the undoped ZnSb case is typical for a semiconductor where ionized impurity scattering dominates at low temperature and acoustic phonon scattering at high temperature. The maximum occurs around 40 K for the undoped case. The temperature where the maximum occurs increases only little for the doped cases to 65-70K. If the only scattering mechanisms had been ionized impurity scattering and acoustical phonon scattering, then the temperature for the maximum would scale as $N_i^{1/3}$, with $N_i$ as the concentration of ionized scattering centers. (See the online supporting information [link.provided.by.the.publisher] for derivation of these scaling laws). Further the maximum value of the mobility would scale as $1/N_i^{1/2}$. If this had applied to the current cases, then from the temperature of the maximum we would have $(70/40)^3 = 5.4$ times as many ionized im-purity scattering centers in [S]=0.02at% doped material as in undoped ZnSb. The ratio between the observed carrier concentration for the [S]=0.02at% doped case and undoped case at the temperature of the max mobility is about $(6\times10^{17}/1\times10^{17}) = 6$, which is a reasonable agreement if it had been assumed that the concentration of ionized impurities was equal to the carrier concentration. On the other hand, from the ratio between the maximum mobility for the [S]=0.02at% doped and undoped cases we would have $(1412/173)^2 = 67$ times as many ionized impurity atoms. This is not the case; meaning that the mobility for the S doped case is reduced much more than can be accounted for by assuming only that the ionized acceptor concentration is equal to the carrier concentration. This indicates that we have additional scattering in the [S]=0.02at% doped case. This could occur by having many neutral impurity scattering centers. That would reduce the mobility. The scattering cross section for neutral impurities is here taken to be temperature independent [29]. Thus, with additional neutral impurities, the temperature for the maximum mobility is not influencing much, but the value of the mobility is decreased. The main difference between the two sulfur cases is that the carrier (hole) concentration is lower by 40 % for the lowest doping concentration and the resistivity is higher by around 40 % between 10 K and 300 K, while at high temperature the resistivity and the carrier concentrations converge for the two S doped cases, which would be expected as the material become intrinsic (n=p) at high temperature.

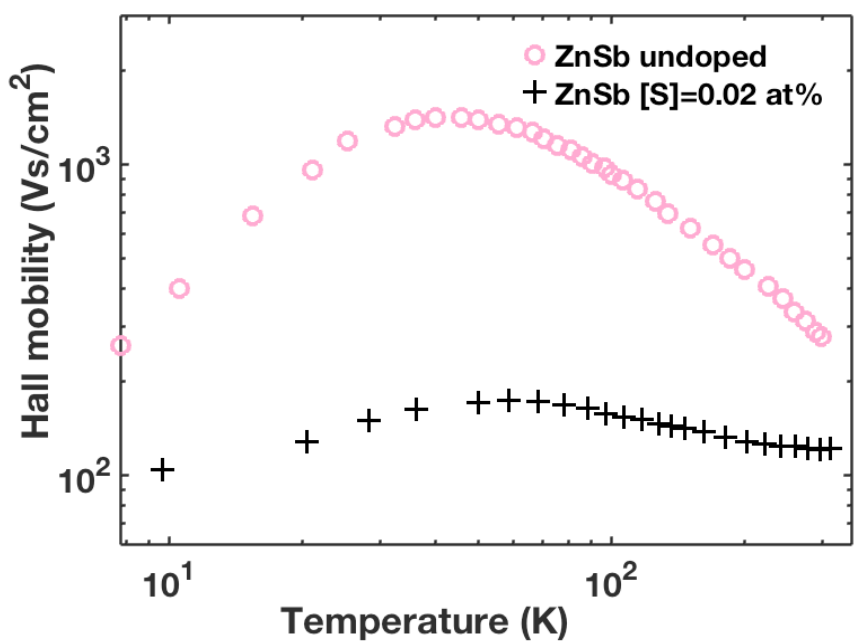

Fig. 8. Hall mobility as a function of measurement temperature for a sample of series B, compared with undoped ZnSb. On this scale the mobility data for [S]=0.1 is very similar to that of [S]=0.02at%.

3.4.4. Explanation of T dependence undoped

A short explanation of the temperature dependence for the undoped case follows: The reason that the Hall carrier concentration decreases with temperature for the lowest temperature (5-35K) is found within the framework of two carrier type conduction; here we have a concentration p in the valence band and $p_1$ in the impurity band with mobilities $\mu$ and $\mu_1$ respectively. At the lowest temperature all the conduction occurs in the impurity band, which is partially filled. However, the resistivity is high because of the low mobility of carriers in the impurity band. When the temperature is increased holes are exited from the impurity band to the valence band. Since the mobility in the valence band is higher than that in the impurity band this causes an increase in conductivity, hence a decrease in resistivity. Similarly the Hall carrier concentration, $p_H$, decreases because of the relative weighting factors by the mobilities [28].

$$p_H = \frac{(\mu p + \mu_1 p_1)^2}{\mu^2 p + \mu_1^2 p_1} \qquad (1)$$

This trend in resistivity changes when the dominant scattering mechanism of carriers in the valence band changes from ionized impurity scattering to acoustic phonon scattering; then the resistivity increases, mainly due to the decrease of mobility with temperature. At high temperatures (>400 K for undoped) the increase in carrier concentration is dominated by generation of hole electron pairs over the band gap giving a strong decrease in resistivity.

3.4.5. Sign of resistivity temperature coefficient

The most striking difference in the temperature dependence of the electrical (transport) properties of the undoped and the sulphur doped samples is the temperature coefficient of the resistivity in the temperature range 100-300 K. For the undoped case it is positive, while for the doped cases it is negative. From writing the resistivity in terms of mobility, $\mu(T)$, and carrier concentration, $p(T)$, we have for the temperature coefficient of resistivity

$$\frac{1}{\rho}\frac{\partial \rho(T)}{\partial T} = \frac{\partial}{\partial T}\log(\rho(T)) = -\frac{\partial}{\partial T}\log(p(T)) - \frac{\partial}{\partial T}\log(\mu(T)) \qquad (2)$$

In the temperature range considered (100-300 K) the sign of the derivative in the first term on the far right hand side is positive for all samples and the derivative in the second term is negative for all samples. Thus the sign of the resistivity coefficient can change if the relative magnitude of the derivatives changes. This is the case for the undoped versus the doped samples. In other words the relative change in the mobility is less for the doped samples because the mobility is strongly influenced by a scattering mechanism that is temperature independent. This is consistent with the indications of a large concentration of neutral scattering centers mentioned above. At high temperatures (400-500 K) the increase in the carrier concentration with temperature is strong and given by an Arrhenius factor with an activation energy equal to half the band gap. The temperature coefficient of resistivity should then be negative.

3.5. Discussion of solid and electrical solubility

3.5.1 Discussion of $[S_{Sb}]$ and solubility

An important observation from a thermoelectric application point of view is the lack of n-type ZnSb with sulfur doping. We can discuss to which degree it might be related to i) low solubility of S in ZnSb, ii) difficulty in formation of $S_{Sb}$, iii) compensation by native defects in ZnSb and iv) introduction of other compensating defects. Regarding the solubility we could not determine the solubility very accurately here. The fact is that no precipitates were observed by SEM and XRD for those samples of series A, which had sulfur concentrations in the melt less than 0.1 at% S. However, the actual solubility could be much less. One of the reasons it could be smaller is related to that the concentration in the samples could likely be smaller than the nominal centration of the melt due to evaporation and redistribution of S. Another effect concerning the solubility is related to the possibility that small clusters of defects incorporating S could form for a sulphur concentration lower than the solubility by an order magnitude, and these clusters are not detected neither by XRD nor SEM. Such effects are observed for example for As in Si where the "electric solubility" is lower than the solid solubility [30]. We suggest that the solubility and electrical solubility of sulfur in ZnSb is much less than 0.02 at% corresponding to $8 \times 10^{18}$ at/cm$^3$ since there is no indications of compensation by S (i.e. the p- concentration does not decrease by increasing [S]). Regarding the difficulty of formation of substitutional sulfur on antimony sites, $S_{Sb}$, we can make comparisons with other cases found in the literature. We will first make a comparison of Te solubility since there have been some reports of n-type behavior (albeit for a small parameter range) for Te in ZnSb [25]. We can apply the Hume-Rothery rules [31] to the comparison of Te solubility with S solubility. The case of S on an Sb site is less favorable than Te on an Sb site because of the larger difference in electronegativity and the larger difference in covalent radius. We can similarly compare S doping in GaAs [32] and S doping in ZnSb. The doping concentration of S in GaAs can be $10^{20}$ cm$^{-3}$ and the covalent radius mismatch is -20 %. For Te in GaAs the electron concentration can also be $10^{20}$ cm$^{-3}$ and the covalent radius mismatch is +20%. For S in ZnSb the mismatch is 40%, while for Te in ZnSb it is only 4%. From these considerations one could expect a higher enthalpy of formation of $S_{Sb}$ than $Te_{Sb}$ and a smaller solid solubility of S on Sb sites. It could also be that other lattice sites are more favorable than the substitutional site, and the solid solubility could be controlled by the thermodynamics of those sites.

3.5.2 Discussion of S lattice locations

The present experiments show that the addition of S to ZnSb is accompanied by the introduction of more defects in the material compared to the undoped case. The hole concentration increases (Fig. 6) and the mobility changes indicates introduction of many uncharged scattering centers as noticed in the discussion above. We suspect S addition is accompanied by an increase in the concentration of Zn vacancies, which would yield increased p-type behavior. Further, at high vacancy concentrations several clusters involving multiple vacancies and possibly sulfur might lead to neutral defects. That Zn vacancies yield p-type behavior has commonly been argued and has also been put into a theoretical framework [16, 33]. Regarding clusters acting as neutral centers, there are several examples of that with other semiconductors, one common case is As in Si [30, 34, 35], another is the $V_{Cu}+Zn_{Cu}$ complex in the photovoltaic material $Cu_2ZnSnS_4$ (CZTS)[36]. We can tentatively forward a simple hypothesis, which would explain parts of the results and lack of n-type behavior. If it was favored to create a cluster with a Zn vacancy and S substituting Sb, i.e. a $S_{Sb}+V_{Zn}$ complex, then the potential donor $S_{Sb}$ was bound in a neutral center and would not contribute any electrons to the conduction band. If the $S_{Sb}$ center attracted Zn vacancies and lowered the ionization energy of the Zn vacancy one could also have more holes than in the undoped case. This would qualitatively explain the increase in neutral scattering centers by introduction of S and it would qualitatively explain the tendency towards increasing hole concentration with increasing S concentration. However, the hy-pothesis lacks any further support, and it would be desirable to calculate theoretically on levels and stability of various clusters.

We may also consider interstitial positions of S, $S_i$, related to the precipitation of ZnS at high concentrations of S in ZnSb. ZnS is an ionic semiconductor where there is charge transfer from Zn to S, making the environment of S having excess electrons. That is not the situation for an ionized S donor on the Sb site. The charge transfer to create the preferred bonds of S (in ZnS) can thus be considered more favorable when S is on an interstitial position than on the Sb site. Similarly the electron transfer from Zn may more easily happen from interstitial Zn than substitutional Zn. Therefor we may hypnotize that a precursor to ZnS precipitation is the formation of $Zn_i^{2+}$-$Sb_i^{2-}$ centers. Therefor an extra Zn vacancy is formed giving an extra hole while the $Zn_i^{2+}$-$Sb_i^{2}$ complex will act as a neutral scattering center, and not contribute free electrons.

4. Summary and conclusions

Sulfur doping of ZnSb has been explored, motivated by a need for n-type ZnSb material for thermoelectric applications. No n-type behavior was found. The S solid solubility in ZnSb is estimated to be lower than 0.1at%, but is probably lower than that, and the solubility of electrically active S is much lower. S doping leads to increased hole concentration and introduces many neutral scattering centers. A tentative simple hypothesis involving clusters of point defects could apparently explain the main trends of the measurements.

Acknowledgement

The help of Ole Bjørn Karlsen in initial sample preparation methodologies is greatly appreciated. This work was supported by the Norwegian Research Council under Contract NFR11-40-6321 (NanoThermo), and the University of Oslo.